\documentclass[sigconf]{acmart}
\usepackage[]{xcolor}
\usepackage{gensymb}
\usepackage{subcaption}
\usepackage{multirow}
\usepackage{array}
\usepackage{soul}
\usepackage{bbm}
\usepackage{algorithm}
\usepackage{algpseudocode}
\usepackage{mathtools}

\DeclareMathOperator*{\argmin}{argmin}

\definecolor{Green}{rgb}{0,0.5,0}

\newcommand{\dd}{\mathsf{d}}
\newcommand{\diag}{\mathrm{diag}}
\usepackage{fnpct}

\newcommand{\oursystem}{PROF}
\AtBeginDocument{%
  \providecommand\BibTeX{{%
    \normalfont B\kern-0.5em{\scshape i\kern-0.25em b}\kern-0.8em\TeX}}}

\setcopyright{acmcopyright}
\copyrightyear{2021}
\acmYear{2021}
\setcopyright{rightsretained}
\acmConference[e-Energy '21]{The Twelfth ACM International Conference on Future Energy Systems}{June 28-July 2, 2021}{Virtual Event, Italy}
\acmBooktitle{The Twelfth ACM International Conference on Future Energy Systems (e-Energy '21), June 28-July 2, 2021, Virtual Event, Italy}\acmDOI{10.1145/3447555.3464874}
\acmISBN{978-1-4503-8333-2/21/06}

\begin{document}

%
\title{Enforcing Policy Feasibility Constraints through Differentiable Projection for Energy Optimization}

\author{Bingqing Chen}
\authornote{These authors contributed equally.}
\orcid{1234-5678-9012}
\affiliation{%
  \institution{Carnegie Mellon University}
  \streetaddress{5000 Forbes Avenue}
  \city{Pittsburgh}
  \state{PA}
  \country{USA}}
\email{bingqinc@andrew.cmu.edu}

\author{Priya Donti}
\authornotemark[1]
\orcid{1234-5678-9012}
\affiliation{%
  \institution{Carnegie Mellon University}
  \streetaddress{5000 Forbes Avenue}
  \city{Pittsburgh}
  \state{PA}
  \country{USA}}
\email{pdonti@cs.cmu.edu}

\author{Kyri Baker}
\orcid{1234-5678-9012}
\affiliation{%
  \institution{University of Colorado, Boulder}
  \streetaddress{}
  \city{Boulder}
  \state{CO}
  \country{USA}}
\email{kyri.baker@colorado.edu}

\author{J. Zico Kolter}
\orcid{1234-5678-9012}
\affiliation{%
  \institution{Carnegie Mellon University}
  \streetaddress{5000 Forbes Avenue}
  \city{Pittsburgh}
  \state{PA}
  \country{USA}}
\email{zkolter@cs.cmu.edu}

\author{Mario Berg\'{e}s}
\orcid{0003-2948-9236}
\affiliation{%
  \institution{Carnegie Mellon University}
  \streetaddress{5000 Forbes Avenue}
  \city{Pittsburgh}
  \state{PA}
  \country{USA}}
\email{marioberges@cmu.edu}


\begin{abstract}
While reinforcement learning (RL) is gaining popularity in energy systems control, its real-world applications are limited due to the fact that the actions from learned policies may not satisfy functional requirements or be feasible for the underlying physical system. In this work, we propose PROjected Feasibility (PROF), a method to enforce convex operational constraints within neural policies. Specifically, we incorporate a differentiable projection layer within a neural network-based policy to enforce that all learned actions are feasible. We then update the policy end-to-end by propagating gradients through this differentiable projection layer, making the policy cognizant of the operational constraints. We demonstrate our method on two applications:~energy-efficient building operation and inverter control. In the building operation setting, we show that \oursystem~maintains thermal comfort requirements while improving energy efficiency by 4\% over state-of-the-art methods. In the inverter control setting, \oursystem~perfectly satisfies voltage constraints on the IEEE 37-bus feeder system, as it learns to curtail as little renewable energy as possible within its  safety set.


\end{abstract}



\maketitle

\section{Introduction}
There has been increasing interest in using learning-based methods such as reinforcement learning (RL) for applications in energy systems control.
However, a fundamental challenge with many of these methods is that they do not respect the physical constraints or functional requirements associated with the systems in which they operate.
Therefore, there have been many calls for embedding safety guarantees into learning-based methods in the context of energy systems applications \cite{zhang2019deep, glavic2019deep, dobbe2020learning}.

    

One common proposal to address this challenge is to provide machine learning methods with ``soft penalties'' to encourage them to learn feasible solutions.
For instance, the authors of \cite{zhang_buildsys2018, chen2019gnu} incentivize their RL-based building HVAC controller to satisfy thermal comfort constraints by adding a constraint violation penalty to the reward function.
While such approaches often involve tuning some weight on the penalty term, recent work has proposed more theoretically-grounded approaches to choosing these weights; for instance, in the setting of approximating AC optimal power flow, the authors of \cite{fioretto2020predicting, chatzos2020high} interpret the weight on their constraint violation penalty as a dual variable, and learn it via primal-dual updates.
\citet{gupta2020deep} adopt a similar approach in an inverter control problem.
However, a challenge with these types of ``soft penalty'' methods in general is that while they \emph{incentivize} feasibility, they do not strictly \emph{enforce} it, which is potentially untenable in safety-critical applications.

\begin{figure*}[h!]
    \centering
    \includegraphics[width=\linewidth]{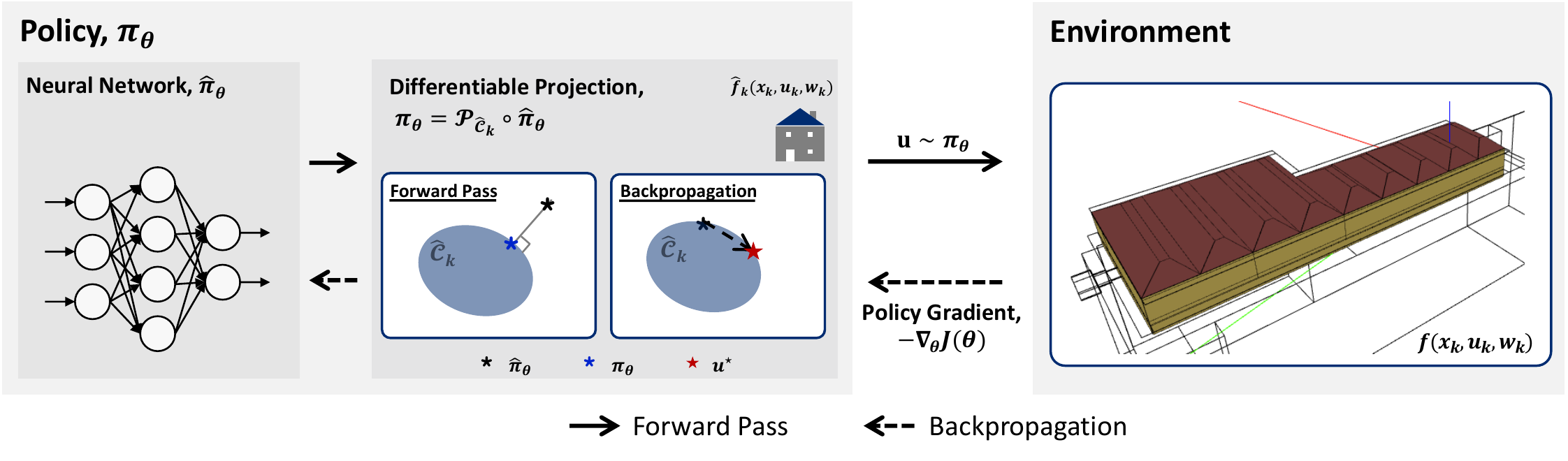}
    \caption{The \oursystem~framework. Our policy consists of a neural network followed by a differentiable projection onto a convexified set of operational constraints, $\hat{\mathcal{C}}_k$ (which is constructed via an approximate model, $\hat{f}_k$, of the environment). The differentiable projection layer enforces the constraints in the forward pass, and induces policy gradients that make the neural network cognizant of the constraints in its learning.
    } 
    \label{fig:framework}
\end{figure*}

Given this limitation, a second class of approaches has aimed to strictly enforce operational constraints.
For instance, in some cases, the outputs of a machine learning algorithm can be clipped post-hoc in order to make them feasible.
However, a challenge is that such post-hoc corrections are not taken into account during the learning process, potentially negatively impacting overall performance.
More recent approaches based in deep learning have therefore aimed to enforce simple classes of constraints in a way that \emph{can} be taken into account during learning; for instance, \citet{zamzam2020learning} train a neural network to approximate AC optimal power flow (OPF), and enforce box constraints on certain variables via sigmoid activations in the last layer of the neural network.
In general, however, existing approaches have only been able to accommodate \emph{simple} sets of constraints, prompting a need for methods that can incorporate broader classes of constraints.


In this work, we propose a method to enforce 
general convex constraints
into RL-based controllers in a way that can be taken into account during the learning process.
In particular, we construct a neural network-based policy that culminates in a \emph{projection} onto a set of constraints characterized by the underlying system.
While the ``true'' constraints associated with the system may be somewhat complex, we observe that simple, approximate physical models are often available for many systems of interest, allowing us to specify convex approximations to the relevant constraints.
The projections onto these (approximate) sets can thus be characterized as convex optimization problems, allowing us to leverage recent developments in differentiable convex optimization \cite{amos2017optnet, agrawal2019differentiable} to train our neural network and projection \emph{end-to-end} using standard RL methods.
The result is a powerful neural policy that can flexibly optimize performance on the true underlying dynamics, while still satisfying the specified constraints.

We demonstrate our PROjected Feasibility approach, \oursystem, on two settings of interest.
Specifically, we explore a building operation setting in which the goal is to reduce energy consumption during the heating season, while ensuring the satisfaction of thermal comfort constraints.
We additionally explore an inverter control setting where the goal is to mitigate curtailment, while satisfying inverter operational constraints and nodal voltage bounds.
In both settings, we find that our controller achieves good performance 
with respect to the control objective, 
while ensuring that relevant operational constraints are satisfied.



To summarize, our key contributions are as follows:
\begin{itemize}
\item \textbf{A framework for incorporating convex constraints.} We propose a projection-based method to flexibly enforce convex constraints within neural policies (as summarized in Figure~\ref{fig:framework}). By examining the gradient fields of the differentiable projection layer, we recommend the incorporation of an auxiliary loss for more robust results. We also show in an ablation study (Section \ref{sec:exp_1results}) that propagating gradients through the differentiable projection layer is indeed conducive to policy learning. 
\item \textbf{Demonstration on building control.} In the building control setting, we show that \oursystem~further improves energy efficiency by 10\% and 4\%, respectively, compared to the best-performing RL agents in \cite{zhang_buildsys2018} and \cite{chen2019gnu}. By using a locally-linear assumption to approximate the building thermodynamics and thereby formulating the constraints as a polytope \cite{zhao2017geometric, chen2020cohort}, we largely maintain the temperature within the deadband, except when the control is saturated.
\item \textbf{Demonstration on inverter control.} In the inverter control setting, \oursystem~satisfies the voltage constraints 100\% of the time over more than half a million time steps (1 week at one second per time step), with a randomly initialized neural network, compared to 22\% over-voltage violations incurred by a Volt/Var control strategy. With respect to the objective of minimizing renewable generation curtailment, \oursystem~performs as well as possible within its conservative safety set after learning safely for a day.
\end{itemize}

\section{Related Work}

Our approach relies on recent developments in implicit neural network layers,
and is thematically similar to several recent works in safe RL.
We briefly discuss these topics, and refer interested readers to \cite{dobbe2020learning, zhang2019deep, glavic2019deep, rolnick2019tackling, drgona2020all} for comprehensive reviews of relevant work in power and energy systems application domains.

\paragraph{Implicit layers} A neural network can be viewed as a composition of functions, or \emph{layers}, with parameters that can be adjusted 
to improve performance on some task.
While many of the layers commonly used within neural networks (e.g., convolutions or sigmoid functions) represent \emph{explicit} functions that provide a direct mapping between inputs and outputs, there has recently been a great deal of interest in expanding the set of commonly-used layers to include those representing \emph{implicit} functions \cite{kolter2020deepimplicit}.
This has included the creation of layers capturing optimization problems \cite{amos2017optnet, djolonga2017differentiable, tschiatschek2018differentiable, wang2019satnet, agrawal2019differentiable, gould2019deep}, physical equations \cite{de2018end, chen2018neural, greydanus2019hamiltonian}, sequence modeling processes \cite{bai2019deep}, and games \cite{ling2018game}. 
In this work, we leverage advances in differentiable optimization in particular, namely by incorporating a differentiable convex optimization layer into our neural policy in order to project proposed control actions onto the feasible set of constraints.

\paragraph{Safe reinforcement learning}
While (deep) RL methods in general lack safety or stability guarantees, there has been recent interest in learning RL-based controllers that attempt to maintain some notion of safety during training and/or inference -- e.g., to satisfy physical constraints or avoid particularly negative outcomes \cite{garcia2015comprehensive}.
These include methods that aim to determine ``safe'' regions of the state space by making smoothness assumptions on the underlying dynamics \cite{wachi18safeexpl, berkenkamp17stability, turchetta16safeexpl, akametalu14reachability}, methods that combine concepts from RL and control theory \cite{morimoto2005robust, luo2014off, pinto2017robust, chang2019neural, han2019h, zhang2020policy, donti2021enforcing},
approaches based on formal verification logics \cite{hunt2020verifiably, hasanbeig2020deep, fulton2019verifiably}, and methods that aim to bound some (discounted) cost function characterizing violations of safety constraints \cite{yang2020projection, taleghan2018efficient, achiam17cpo, altman1999constrained}.
While the particular notion of ``safety'' considered varies between settings, relevantly to the present work, several of these prior works employ some form of differentiable projection within the loop of deep RL.
For instance, within the context of constrained Markov decision processes (C-MDPs), \citet{yang2020projection} project neural network-based policies onto a linearly-constrained set of policies with bounded cumulative discounted cost.
In the context of asymptotic stability, \citet{donti2021enforcing} project the actions output by their controller onto a convex set of actions satisfying stability specifications obtained via robust control.
In the setting of robotic motion planning, \citet{pham2018optlayer} project actions onto a linear set of robotic operational constraints, and apply separate updates to the neural network based on both pre-projection and post-projection actions.
Similarly to this prior work, our approach employs differentiable projections within a neural network policy to enforce operational constraints over some planning horizon.

\section{Preliminaries}\label{sec:background}

We now present background on technical concepts used by \oursystem, namely reinforcement learning and differentiable projection layers.

\subsection{Reinforcement Learning} \label{sec: RL}
The goal of RL is to learn an optimal control policy through direct interaction with the environment. The problem is usually formulated as a Markov decision process (MDP). At each time step $k$, the agent selects an action $u_k$ given the current state $x_k$, using its policy $\pi_\theta$ (Equation~\ref{eq:policy}). 
In many modern RL techniques, the policy is commonly represented by a neural network parameterized by $\theta$. When the agent takes the action $u$, the state transitions to $x'$ based on the system dynamics $f$ (Equation~\ref{eq:dynamics}), and the agent receives a reward $r_k$ (or equivalently, incurs a cost $c_k = -r_k$). 
\begin{align} 
  u &\sim \pi_{\theta}(u_k | x_k), \label{eq:policy} \\
  x' &\sim f(x_k,u_k). \label{eq:dynamics}
\end{align}
RL algorithms optimize for a policy that maximizes the expected cumulative reward, or equivalently, minimizes the expected cumulative cost, where $\gamma$ is a temporal discount factor:
\begin{equation}
 \label{eq:objective}
\theta^\star = \arg\max_{\theta}{\mathbb{E}_{\pi_\theta}} \left[\sum_{l=0}^\infty \gamma^l r_{k+l}\right] = \arg\min_{\theta}{\mathbb{E}_{\pi_\theta}} \left[\sum_{l=0}^\infty \gamma^l c_{k+l}\right].
\end{equation}
To simplify notation, we will denote the expected cumulative cost as $J(\theta)$, i.e., 
\begin{equation}
J(\theta) = \mathbb{E}_{\pi_\theta} \left[\sum_{l=0}^\infty \gamma^l c_{k+l}\right].
\end{equation}

There are three general approaches to RL, namely value-based methods, policy gradient methods, and actor-critic methods.
Value-based methods, e.g., Q-learning and its variants, update the value function of state-action pairs using the Bellman equation and take the action that maximizes the value of an action selection policy (the Q function) through exploration. Policy gradient methods, e.g., Proximal Policy Optimization (PPO) \cite{schulman2017proximal}, directly search for an optimal policy $\pi_{\theta}^\star$ using estimates of policy gradients. Denoting the policy gradient as $g:= \nabla_\theta J(\theta)$, the core idea of  policy gradient algorithms is that they update $\theta$ based on an estimate, $\hat{g}$, of the gradient, i.e.,
\begin{equation}\label{eq:pg}
    \theta \xleftarrow{} \theta - \alpha \hat{g}
\end{equation}
for some learning rate $\alpha$. Different algorithms vary in how they obtain $\hat{g}$. For instance, the learning objective for PPO, which we use in our building control experiment (Section~\ref{sec:exp-1}), is given by the following equation, where $\hat{A}_t$ is the generalized advantage estimate that can be estimated via any of the estimators in \cite{schulman2015high}:
\begin{equation}\label{eq:PPO}
\begin{aligned}
J_{\text{PPO}}(\theta)&=\hat{\mathbb{E}}_k\left[\min(w_k(\theta)\hat{A}_k, \text{clip}(w_k(\theta), 1-\epsilon, 1+\epsilon)\hat{A}_k) \right],\\
    w_k(\theta) &= \frac{\pi_\theta(u_k|x_k)}{\pi_{\theta_{old}}(u_k|x_k)},
\end{aligned}
\end{equation}
and the estimate $\hat{g}$ is constructed based on this learning objective.
Actor-critic methods, e.g., Advantage Actor-Critic (A2C), are hybrids of the value-based and policy gradient approaches, using a policy network to select actions (the actor) and a value network to evaluate the action (the critic).


\subsection{Differentiable Projection Layers}
\label{sec:diff-proj-layers}

As previously described, a neural network is a composition of parameterized functions (\emph{layers}) whose parameters are adjusted during training via \emph{backpropagation} (a class of gradient-based methods).
Any function can be incorporated into a neural network as a layer provided that it satisfies two main conditions.
The first condition is that it must have a \emph{forward procedure} to map from inputs to outputs (i.e., do \emph{inference}).
The second is that it must have a \emph{backwards procedure} to compute gradients of the outputs with respect to the inputs and function parameters, in order to enable backpropagation.

With that in mind, consider the $L_2$-norm projection $\mathcal{P}_{\mathcal{C}}: \mathbb{R}^n \to \mathcal{C}$ that maps from some point in $\hat{u} \in \mathbb{R}^n$ to its closest point in some constraint set $\mathcal{C} \subseteq \mathbb{R}^n$ as follows:
\begin{equation}
    \mathcal{P}_{\mathcal{C}}(\hat{u}) = \argmin_{u \in \mathcal{C}} \frac{1}{2} \|u - \hat{u} \|_2^2.
\label{eq:generic-proj}
\end{equation}
In cases where $\mathcal{C}$ is convex, Equation~\ref{eq:generic-proj} is a convex optimization problem.
The forward procedure of this operation can then be implemented by simply solving the optimization problem, e.g., using standard convex optimization solvers.
Perhaps less evidently, it is also possible to construct a backwards procedure for this problem by using the implicit function theorem \cite{krantz2012implicit}, as described in previous work (e.g., \cite{amos2017optnet, agrawal2019differentiable}).

As an example, consider the case where $\mathcal{C}$ characterizes linear constraints, i.e., $\mathcal{C} \equiv \{ u : A u = b, G u \leq h \}$ for some $A \in \mathbb{R}^{n_{\text{eq}} \times n}$, $b \in \mathbb{R}^{n_{\text{eq}}}$, $G \in \mathbb{R}^{n_{\text{ineq}} \times n}$, and $h \in \mathbb{R}^{n_{\text{ineq}}}$.
It is then possible to efficiently compute gradients through Equation~\ref{eq:generic-proj} by implicitly differentiating through its KKT conditions, i.e., conditions that are necessary and sufficient to describe its optimal solutions.
In particular, as described in \cite{amos2017optnet}, the KKT conditions for stationarity, primal feasibility, and complementary slackness for this case are given by
\begin{equation}
\begin{split}
u^\star - \hat{u} +A^T\nu^\star+G^T\lambda^\star &= 0 \\
Au^\star-b &= 0 \\
\diag(\lambda^\star)(Gu^\star-h) &= 0,
\end{split}
\label{eq:kkt}
\end{equation}
where $u^\star$, $\lambda^\star$, and $\nu^\star$ are the optimal primal and dual solutions.
By the implicit function theorem, we can then take derivatives through these conditions at the optimum in order to obtain relevant gradients.
Specifically, the total differentials of these KKT conditions are given by
\begin{equation}
\begin{split}
\dd u - \dd \hat{u} + \dd A^T \nu^\star +
A^T \dd \nu + \dd G^T
\lambda^\star + G^T \dd \lambda & = 0 \\
\dd A u^\star + A \dd u - \dd b & = 0 \\
\diag(Gu^\star -h)\dd \lambda + \diag(\lambda^\star)(\dd G u^\star  + G \dd z  - \dd h)
& = 0.
\end{split}
\end{equation}
As described in \cite{amos2017optnet}, these equations can then be rearranged to solve for the Jacobians of any of the solution variables $u^\star, \lambda^\star, \nu^\star$ with respect to any of the problem parameters $\hat{u}, A, b, G, h$ (or, in practice, to solve directly for these Jacobians' left matrix-vector product with some backward pass vector, in order to reduce space complexity).

While the above example is for the case of a linearly-constrained projection operation, these kinds of gradients can be computed for convex projection problems in general.
For instance, \citet{donti2021enforcing} compute gradients through a projection onto a second order cone by differentiating through the fixed point equations of their solver, and \citet{agrawal2019differentiable} provide a method and library for differentiable disciplined convex programs.
A key benefit of using these kinds of projection layers for constraint enforcement is that they 
allow gradients through the enforcement procedure to flow back to the neural network, thereby informing the parameter updates of this network during training. 

\section{Approach}

We now describe \oursystem, which incorporates differentiable projections onto convex(ified) sets of operational constraints within a neural policy.

\subsection{Problem Formulation}
Consider a discrete-time dynamical system
\begin{equation}
    x_{k+1} = f(x_k, u_k, w_k), 
\label{eq:general-dyn}
\end{equation}
where $x_k \in \mathbb{R}^s$ is the state at time $k$, $u_k \in \mathbb{R}^a$ is the control input, $w_k \in \mathbb{R}^d$ is an uncontrollable disturbance (which we assume to be observable), and $f : \mathbb{R}^s \times \mathbb{R}^a \times \mathbb{R}^d \to \mathbb{R}^s$ denotes the system dynamics. 
Letting $\mathcal{X}_k$ and $\mathcal{U}_k$ denote the allowable state and action space, respectively, we can define the set of all feasible actions over the planning horizon $T$ as $\mathcal{C}_k$, where
\begin{equation}\label{eq:def_C} 
    \mathcal{C}_k= \left\{u_{k:k+T-1}\;
    \Bigg\vert \begin{array}{l}
    x_{i+1} =  f(x_i, u_i, w_i),\\
     \;x_{i} \in \mathcal{X}_{i}, \;  u_i \in \mathcal{U}_i
    \end{array}
    \;\; \forall i \in \{k,...,k+T-1\}
    \right\}.
\end{equation}
%
Our goal is then to learn a policy that optimizes the control objective, $J$, while enforcing the operational constraints. 
To simplify notation, we denote $\mathbf{u} = u_{k:k+T-1}$.  
In the case of a deterministic policy, i.e., $\mathbf{u} = \pi_{\theta}$, the learning problem is simply
\begin{equation}\label{eq:prob-det}
\begin{aligned}
\min_{\theta} \;\; J(\theta)
\quad \textrm{s.t.} \;\; \pi_\theta \in \mathcal{C}_k .
\end{aligned}
\end{equation}
In the case of a stochastic policy, e.g. $\mathbf{u} \sim \mathcal{N}(\boldsymbol{\mu}, \diag(\boldsymbol{\sigma}^2)), \; [\boldsymbol{\mu}, \boldsymbol{\sigma}] = \pi_{\theta}(x_k) $, we can write the problem as 
\begin{equation}\label{eq:prob-stoch}
\begin{aligned}
\min_{\theta} \;\; J(\theta)
\quad \textrm{s.t.} \;\; \mathbf{u}, \boldsymbol{\mu} \in \mathcal{C}_k.
\end{aligned}
\end{equation}
In this case, it is necessary to sample actions around $\boldsymbol{\mu}$ in order to estimate policy gradients. At the same time the actions sampled from $\pi_\theta$ might fall outside of $\mathcal{C}_k$. Thus, we enforce that both $\boldsymbol{\mu}$ and the sample action $\mathbf{u}$ satisfy the constraints.  

\subsection{Approximate Convex Constraints}


In practice, there are two key challenges inherent in solving Equations~\ref{eq:prob-det}--\ref{eq:prob-stoch} as written.
The first is that the disturbances $w_i$ are not known ahead of time, meaning that the optimization problem must be solved under uncertainty.
One approach to addressing this, from the field of robust control \cite{zhou1998essentials}, involves constructing an uncertainty set over the disturbance, and then optimizing for worst-case or expected cost under this uncertainty set.  
Here, we simply assume a predictive model of the disturbances is available.
(By re-planning frequently, we observe that the prediction errors have limited empirical impact on performance in the two applications we study.)
We will use the notation $\hat{w}_k$ to denote our forecast of the disturbance if $k$ is a future time step, and the true value of the disturbance if $k$ is the present or a prior time step.

The second challenge pertains to the form of the set $\mathcal{C}_k$, which may be poorly structured or otherwise difficult to optimize over.
In particular, our framework relies on obtaining convex approximations to the constraints in order to enable differentiable projections (see Section~\ref{sec:diff-proj-layers}). 
Fortunately, for many energy systems applications, some approximate model $\hat{f}_k$ is often available based on domain knowledge that allows $\mathcal{C}_k$ to be approximated as a convex set, despite the complex nature of the true dynamical system.

Thus, letting $\hat{f}_i$ denote our approximations of the dynamics and $\hat{w}_i$ denote the (forecast or known) disturbance at each $i = k, \ldots, k+T-1$, we define our approximate convex constraint set as
\begin{equation}\label{eq:def_Chat} 
    \hat{\mathcal{C}}_k= \left\{u_{k:k+T-1}\;
    \Bigg\vert \begin{array}{l}
    x_{i+1} =  \hat{f}_i(x_i, u_i, \hat{w}_i),\\
     \;x_{i} \in \mathcal{X}_{i}, \;  u_i \in \mathcal{U}_i
    \end{array}
    \;\; \forall i \in \{k,...,k+T-1\}
    \right\}.
\end{equation}
We note that $f$ and $w$ are approximated \emph{solely} for the purposes of constructing approximate constraint sets, and are not used otherwise during training and inference (i.e., our neural policy interacts with the \emph{true} dynamics and disturbances during training and inference).

    \begin{algorithm}[t!]
		\caption{\oursystem}
		\begin{algorithmic}[1]
		    \Procedure{main}{env, $J$} \hspace{3pt} \emph{// input: environment, control objective}
    		    \State \textbf{init} neural network $\hat{\pi}_\theta$, replay memory $\mathcal{M}$  
    		    \State \textbf{specify} RL algorithm $\mathcal{A}$, batch size $M$, update interval $K$
    		    \State \textbf{specify} planning horizon $T$
    		    \State \emph{// online execution}
    		    \For{$k = 1, \ldots$}
    		    \State \textbf{observe} state $x_k$ 
    		    \State \textbf{predict} future disturbances  $\hat{w}_{k:k+T-1}$
    		    \State \textbf{construct} constraint set $\hat{\mathcal{C}}_k$, policy $\pi_\theta = \mathcal{P}_{\hat{\mathcal{C}}_{k}} \circ \hat{\pi}_\theta$
    		    \State \textbf{compute} $u_{k}$ = \Call{inference}{${\pi}_\theta$,  $x_k$, $T$}
    		    \State \textbf{execute action} \Call{env.step}{$u_k$}
    		    \State \textbf{save} \Call{memory.append}{$x_k$, $u_k$, $\hat{w}_{k:k+T-1}$}
    		    \State \emph{// update policy every $K$ time steps}
    		    \If{$\bmod(k, K) = 0$}
    		        \State $\hat{\pi}_\theta$ = \Call{train}{$\hat{\pi}_\theta$,  $J$, $\mathcal{M}$, $\mathcal{A}$} 
    		    \EndIf
    		    \EndFor
		    \EndProcedure
		   \State
		   \Procedure{inference}{$\pi_\theta$,  $x_k$, $T$}
    		    \State \emph{// input: neural policy, current state, planning horizon}
    		    
    		    \State  \textbf{select action} $u_{k:k+T-1} \sim \pi_\theta$
    		  
    		    \emph{// only return the current action; replan at each time step}
    		    \State \textbf{return} $u_{k}$ 
    		  \EndProcedure
		    \State
		   \Procedure{train}{$\hat{\pi}_\theta$,  $J$, $\mathcal{M}$, $\mathcal{A}$} \hspace{3pt} \State\emph{// input: neural policy,  objective, replay memory, RL algorithm} 
		        \State \textbf{init} $\mathcal{L}(\theta)$ = 0
		        \For{$i = 1, \ldots, M$}
		        \State \textbf{sample} $x,u, w\sim \mathcal{M}$
		      \State \textbf{construct} constraint set $\hat{\mathcal{C}}_k$, policy $\pi_\theta = \mathcal{P}_{\hat{\mathcal{C}}_{k}} \circ \hat{\pi}_\theta$
    
    		    \State \textbf{compute} training loss $$\mathcal{L}(\theta) \mathbin{{+}{=}} J(\theta) + \lambda \|\pi_\theta(x) - \hat{\pi}_\theta(x)\|_2^2$$
    		    \EndFor
    		    \State \textbf{train} $\hat{\pi}_\theta$ via $\mathcal{A}$  to minimize $\mathcal{L}$ 
    		    \State \textbf{return} $\hat{\pi}_\theta$
		    \EndProcedure
		\end{algorithmic}
		\label{alg:main-alg}
	\end{algorithm}

\subsection{Policy Optimization}
Let $\hat{\pi}_\theta$ be any (e.g., fully-connected or recurrent) neural network parameterized by $\theta$. 
Our policy entails passing the output from the neural network to the differentiable projection layer $\mathcal{P}_{\hat{\mathcal{C}}_k}$ characterized by the approximate constraints, 
which enforces that the resultant action is feasible with respect to these constraints. The overall (differentiable) neural policy is then given by 
\begin{equation}
    \pi_\theta(x_k) = \mathcal{P}_{\hat{\mathcal{C}_k}} \circ \hat{\pi}_\theta(x_k).\footnote{We use the notation $f\circ g(x) \coloneqq f(g(x))$ to denote function composition.} 
\label{eq:general-policy}
\end{equation}

The key benefit of embedding a differentiable projection into our policy is that it enforces constraints in a way that is visible to the neural network during learning. In this work, we implement the differentiable projection using the \texttt{cvxpylayers} library \cite{agrawal2019differentiable}. 

We construct the following loss function, which is a weighted sum of the control objective $J$ and an auxiliary loss term to be explained shortly in this section. $\lambda>0$ is a hyperparameter.  
\begin{equation}
    \mathcal{L}(\theta, x_k) = J(\theta) + \lambda \| \pi_\theta(x_k) - \hat{\pi}_\theta(x_k) \|_2^2.
    \label{eq:aug-cost}
\end{equation}
We then train our policy (Equation~\ref{eq:general-policy}) to minimize this cost using standard approaches in deep reinforcement learning. The full algorithm is presented in Algorithm~\ref{alg:main-alg}.

\begin{figure}
    \centering
    \includegraphics[width = \linewidth]{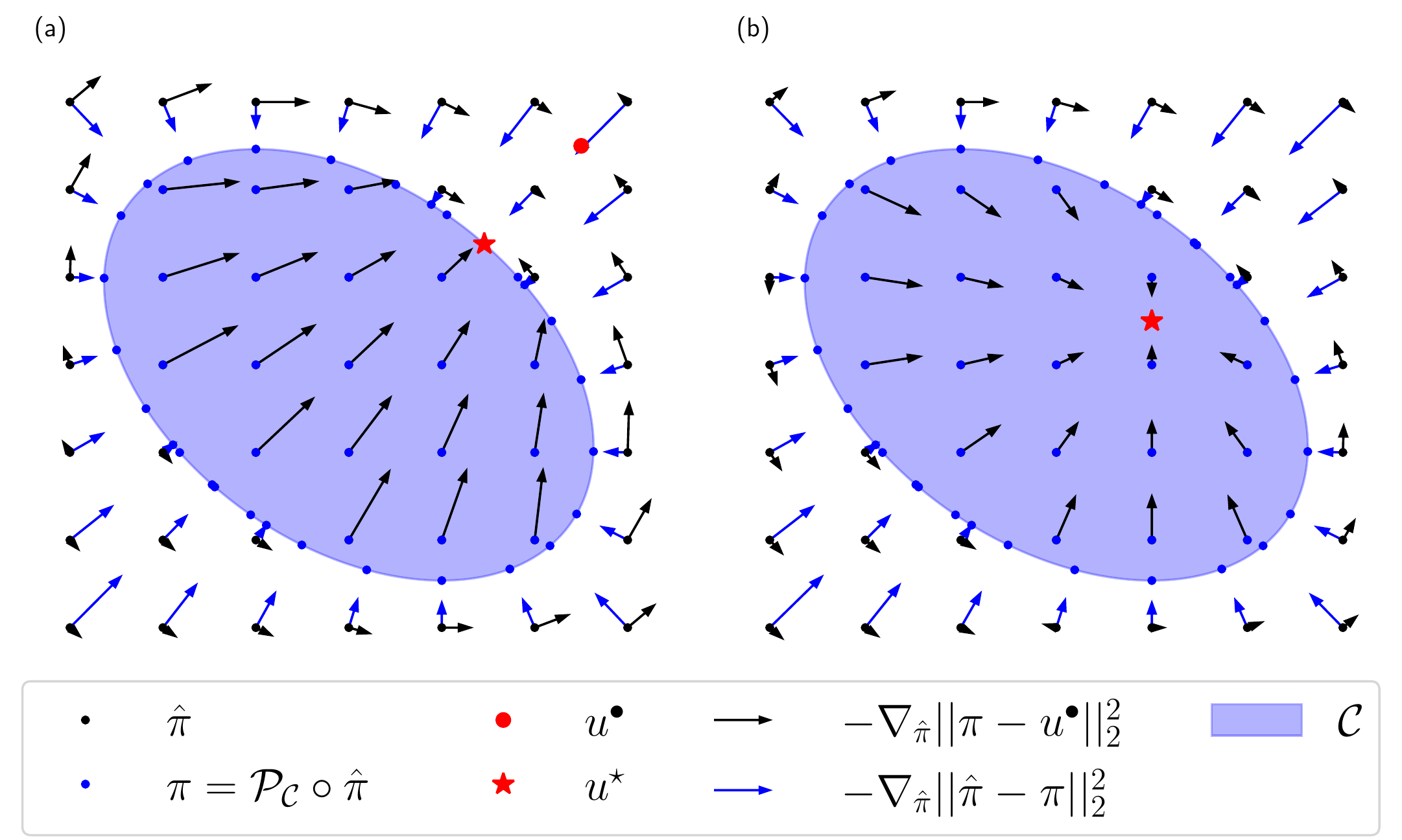}
    \caption{Illustrative example of gradients from the differentiable projection layer. $u^{\bullet}$ and $u^{\star}$ denote unique optimal actions minimizing some convex control objective $J$ in the unconstrained and constrained settings, respectively; $\nabla_{\hat{\pi}}\|\pi - u^{\bullet}\|_2^2$ is thus a proxy for $\nabla_{\hat{\pi}}J$. (a) $u^{\bullet}\not \in \mathcal{C}$. The gradients $\nabla_{\hat{\pi}}J$ point towards $u^{\star}$ as desired, such that $\pi = \mathcal{P}_{\hat{\mathcal{C}}} \circ \hat{\pi}$ will reach this optimal point. (b) $u^{\bullet} = u^{\star}$ on the interior of $\mathcal{C}$. The gradients $\nabla_{\hat{\pi}}J$ do not cause $\hat{\pi}$ (or its projection) to update towards the interior. Adding a weighted auxiliary loss term, e.g., $\|\pi-\hat{\pi}\|$, can help direct updates towards the interior.}
    \label{fig:gradients}
\end{figure}

\subsubsection{Visualization of gradient fields.} To provide more intuition on the differentiable projection layer and our cost function, we visualize the gradient fields in a hypothetical example with a deterministic policy and a planning horizon of $T=1$. 
Specifically, for the purposes of illustration, let $u^{\bullet}$ and $u^{\star}$ denote unique optimal actions minimizing some convex control cost $J$ in the unconstrained and constrained settings, respectively:
\begin{equation*}\label{eq:u_star}
\begin{aligned}
u^{\bullet} \sim \pi_{\theta^{\bullet}};\quad &\theta^{\bullet} = \arg\min_{\theta} \;\; J(\theta)\\
u^{\star}\sim \pi_{\theta^{\star}};\quad
&\theta^{\star} = \arg\min_{\theta} \;\; J(\theta)
\quad \textrm{s.t.} \;\; u \in \mathcal{C}_k.
\end{aligned}
\end{equation*}
In Figure~\ref{fig:gradients}, we then plot the gradient fields in two cases: (a) $u^{\bullet} \not \in \mathcal{C}_k $, and (b) $u^{\bullet} \in \mathcal{C}_k$.
Note that $u^{\bullet}$ and $u^{\star}$ are assumed to be known here for illustrative purposes only, and are not known during training. 


In particular, we plot the gradients (black arrows) of $\|u^{\bullet} - \mathcal{P}_{\mathcal{C}_k} \circ \hat{\pi}\|_2^2$ with respect to the output of the neural network $\hat{\pi}$. 
These indicate the direction in which the neural network would be incentivized to update in order to minimize the system cost.
If no differentiable projection were embedded within the policy, all the gradients would point towards $u^{\bullet}$ without regard for the constraints. Instead, in the case of $u^{\bullet} \not \in \mathcal{C}_k $ (Figure~\ref{fig:gradients}a), the gradients through the differentiable projection layer point towards $u^\star$ instead of $u^\bullet$.  
More specifically, if $\hat{\pi}_\theta(x_k) \in \mathcal{C}_k$, then the projection layer is simply the identity, and the gradients point directly towards $u^\star$; otherwise, the gradients point along the boundary of $\mathcal{C}_k$ in the direction of $u^{\star}$.

This case is of particular interest, as in many practical applications some operational constraint will be binding.
As a concrete example, the ultimate energy-saving strategy for building operations is to keep all mechanical systems off (i.e., $u^{\bullet} = 0$), which obviously violates occupants' comfort requirements and is outside the set of allowable actions (i.e., $u^{\bullet}\not \in \mathcal{C}_k$). Thus, the problem is to find a policy that uses the mechanical system as little as possible without violating comfort requirements.
Given the common case where the control objective is convex, this then lies on the boundary of the constraint set (i.e., $u^\star = \mathcal{P}_{\mathcal{C}_k} \circ u^\bullet$).

We also depict the case where the solution of the unconstrained problem already satisfies the constraints, i.e., $u^{\bullet}=u^\star\in C_k$ (Figure~\ref{fig:gradients}b).  
If this is generally the case for a particular application, we note that a constraint enforcement approach (ours or otherwise) is likely not needed, and indeed utilizing gradients through the projection layer may actually degrade performance.
Specifically, if $\hat{\pi}_\theta(x_k) \not \in \mathcal{C}_k$, the gradients do not point towards the interior of the constraint set, meaning that $\pi_\theta(x_k) = \mathcal{P}_{\mathcal{C}_k} \circ \hat{\pi}_\theta(x_k)$ will lie on the boundary of the constraints despite the optimal solution being in the interior.
This can be amended by augmenting the loss function with a (weighted) auxiliary term such as $\| \pi_\theta(x_k) - \hat{\pi}_\theta(x_k) \|_2^2$ whose gradients (blue arrows) point towards the interior. 

It may not be known a priori whether or not $u^{\bullet}$ is in the constraint set in general or at any given time, except when domain experts are fully clear on the structure of the solutions for specific applications. 
In particular, $\mathcal{C}_k$ is time-varying, making it difficult to know for sure whether or not the constraints will indeed be binding at any given time. For robustness, we therefore recommend incorporating the auxiliary loss $\| \pi_\theta(x_k) - \hat{\pi}_\theta(x_k) \|_2^2$ within the RL training cost, unless it is known from domain knowledge that the constraints will certainly be active.
As such, we formulate the training cost function as previously given in Equation~\ref{eq:aug-cost}.


\section{Experiment 1: Energy-efficient Building Operation}\label{sec:exp-1}
There is significant potential to save energy through more efficient building operation. Buildings account for about 40\% of the total energy consumption in the United States, 
and it is estimated that up to 30\% of that energy usage may be reduced through advanced sensing and control strategies \cite{fernandez2017impacts}. However, this potential is largely untapped, as the heterogeneous nature of building environments limits the ability of control strategies developed for one building to scale to others \cite{chen2019gnu}. RL can address this challenge by adapting to individual buildings by directly interacting with the environment. 

The most important constraint in building operation is to maintain a satisfactory level of comfort for occupants, while minimizing energy consumption. It is common in the RL-based building control literature to penalize thermal comfort violations \cite{zhang_buildsys2018, chen2019gnu}, which incentivizes but does not guarantee the satisfaction of these comfort requirements. In comparison, our proposed neural policy can largely maintain temperature within the specified comfortable range, except when the control is saturated.

We evaluate our policy in the same simulation testbed as \cite{zhang_buildsys2018, chen2019gnu}, following the same experimental setup as \cite{chen2019gnu}. Specifically, we first pre-train the neural policy by imitating a proportional-controller (P-controller). We then evaluate and further train our agent in the simulation environment, using a different sequence of weather data. 

\subsection{Problem Description}

\begin{figure}
	\centering	
	\begin{subfigure}{0.28\textwidth} 
	\includegraphics[width=\textwidth]{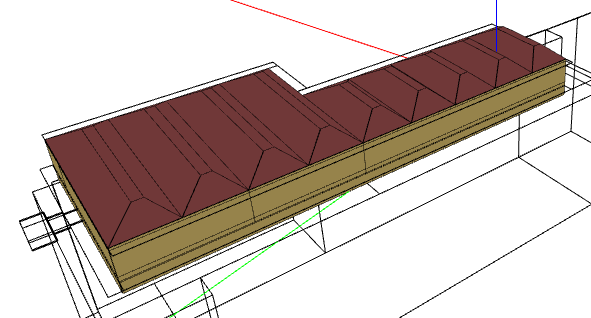}
	\caption{Geometric view
	}
	\label{fig:IW_Geometry}
	\end{subfigure}
	\begin{subfigure}{0.18\textwidth} 
	\includegraphics[width=\textwidth]{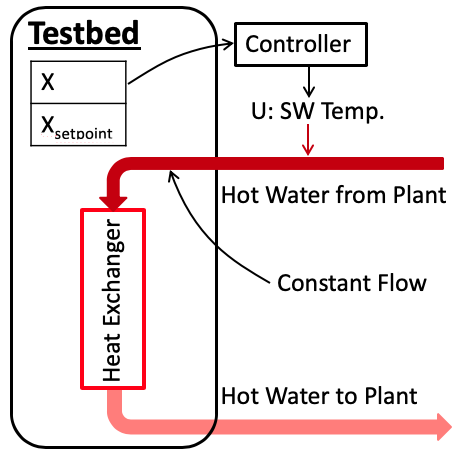}
	\caption{System schematic} 
	\label{fig:IW_schematic}
	\end{subfigure}
	\vspace{-0.3cm}
	\caption{Building simulation testbed (reproduced from \cite{chen2019gnu}).} 
\end{figure}

\paragraph{Simulation testbed} We utilize an EnergyPlus (E+) model of a 600m$^2$ multi-functional space (Figure \ref{fig:IW_Geometry}), based on
\makeatletter
\if@ACM@anonymous
\textit{a particular university building.}
\else
the Intelligent Workplace (IW) on Carnegie Mellon University (CMU) campus, located in Pittsburgh, PA, USA.
\fi
\makeatother
The system of interest is the water-based radiant heating system, of which a schematic is provided in Figure \ref{fig:IW_schematic}.
In this experiment, we control the \textit{supply water temperature} so as to maintain the state variable, i.e., the \textit{zone temperature}, within a comfortable range during the heating season. In the existing control, the supply water (SW) is maintained at a constant flow rate, and its temperature is managed by a P-controller. For more information on the simulation testbed, refer to \cite{zhang_buildsys2018}.

\paragraph{Approximate system model} 
We approximate the environment as a linear system as follows:
\begin{equation} \label{eq:linear_IW}
x_{k+1} \approx \hat{f}(x_k, u_k, w_k)= Ax_k + B_u u_k + B_d w_k,
\end{equation} where $x_{k}$  represents the \textit{zone temperature} and $u_{k}$ represents the \textit{supply water temperature}. $w_{k}$ includes distributions from weather and occupancy. 
While building thermodynamics are fundamentally nonlinear, the locally-linear assumption works well for many control inputs \cite{privara2013building}. 
We identify the approximate model parameters $A$, $B_u$, and $B_d$ with prediction error minimization \cite{privara2013building} on the same data used to pre-train the RL agent (see Section~\ref{sec:building-impl}). 
The root mean squared error (RMSE) of this model on a unseen test set is 0.14\textsuperscript{o}C.


\paragraph{Objective} Since our goal is to minimize energy consumption, we define the control cost at each time step as the agent's control action, i.e. \textit{supply water temperature}, which is linearly proportional to the heating demand, i.e., 
$c_k = u_k.$

In contrast to the objectives in \cite{zhang_buildsys2018, chen2019gnu}, which are defined as weighted sum of energy cost and some penalty on thermal comfort violations,  we consider the thermal comfort requirement as hard constraints, in the form of Equation~\ref{eq:prob-stoch}. 

\paragraph{Constraints} To maintain a satisfactory comfort level, we require the zone temperature to be within a deadband $\mathcal{X}= 
\{x\;|\;21.9\textsuperscript{o}C\leq x \leq 25.5\textsuperscript{o}C\}$ when the building is occupied, based on the building code requirement of 10\% Predicted Percentage of Dissatisfied (PPD) \cite{fanger1986thermal}.
We allow for a wider temperature range during unoccupied hours. For the action, the allowable range of supply water temperature for the physical system is $\mathcal{U}= 
\{u\;|\;20\textsuperscript{o}C\leq u \leq 65\textsuperscript{o}C\}$. 

While it may appear from this description that we have only simple box constraints on both the state and action, we highlight the fact that actions are coupled over time through the building thermodynamics \cite{zhao2017geometric}. More concretely, a future state depends on all past actions. Thus, a box constraint on $x_{k+l+1}$ is in fact a constraint on $u_{k:k+l}$. 
In this case, assuming $\hat{f}$ to be a linear system, $\hat{\mathcal{C}}_k$ is then a set of linear inequalities, which can be geometrically interpreted as a polytope.\footnote{A polytope can be characterized as a set $\mathcal{S}=\{x\in \mathbb{R}^n|Ax\leq b\}$.} We refer interested readers to \cite{chen2020cohort, zhao2017geometric} for more details on this formulation. In fact, it was experimentally demonstrated in \cite{chen2020cohort} that projecting actions onto the polytope constructed with an approximate linear model was sufficient to maintain temperature within the deadband in a real-world residential household (though \cite{chen2020cohort} did not then differentiate through this projection).

\paragraph{Control time step} The  EnergyPlus model has a 5-minute simulation time step. Following \cite{zhang_buildsys2018, chen2019gnu}, we use a 15-min control time step (i.e., each action is repeated 3 times) and a planning horizon of $T = 12$ (i.e., a 3 hour look-ahead). 

\subsection{Implementation Details}
\label{sec:building-impl}
\paragraph{Offline pre-training} We pre-train a long short-term memory (LSTM) recurrent policy (without a subsequent projection) by imitating a P-controller operating under the Typical Meteorological Year 3 (TMY3) ~\cite{wilcox2008users} weather sequence, from Jan. 1 to Mar. 31. We min-max normalize all of the state, action, and disturbance, and use a learning rate of $10^{-3}$. Specifically, we use the pre-trained weights after training on the expert demonstrations for 20 epochs following the same procedures as \cite{chen2020towards}. We refer readers to \cite{chen2020towards} for more details on the neural network architecture, training procedures, loss, and performance evaluation. 

\paragraph{Online policy learning} We optimize the policy with PPO \cite{schulman2017proximal} over the weather sequence in 2017 from Jan. 1 to Mar. 31. We use $\lambda=10$ (see Equation~\ref{eq:aug-cost}), a learning rate of $5 \times 10^{-4}$, and RMSprop \cite{tieleman2012lecture} as the optimizer\footnote{The code is available at \url{https://github.com/INFERLab/PROF}.}. We update the policy every four days, by iterating over those samples for 8 epochs with a batch size of 32. For hyperparameters, we use a temporal discount rate of $\gamma$ = 0.9, $\epsilon$ = 0.2 (see Equation~\ref{eq:PPO}), and a Gaussian policy (see Equation~\ref{eq:prob-stoch}) with $\sigma$ linearly decreased from 0.1 to 0.01. 
\subsection{Results}\label{sec:exp_1results}

\begin{figure}
	\centering	\begin{subfigure}{\linewidth} 
	\includegraphics[width=\textwidth]{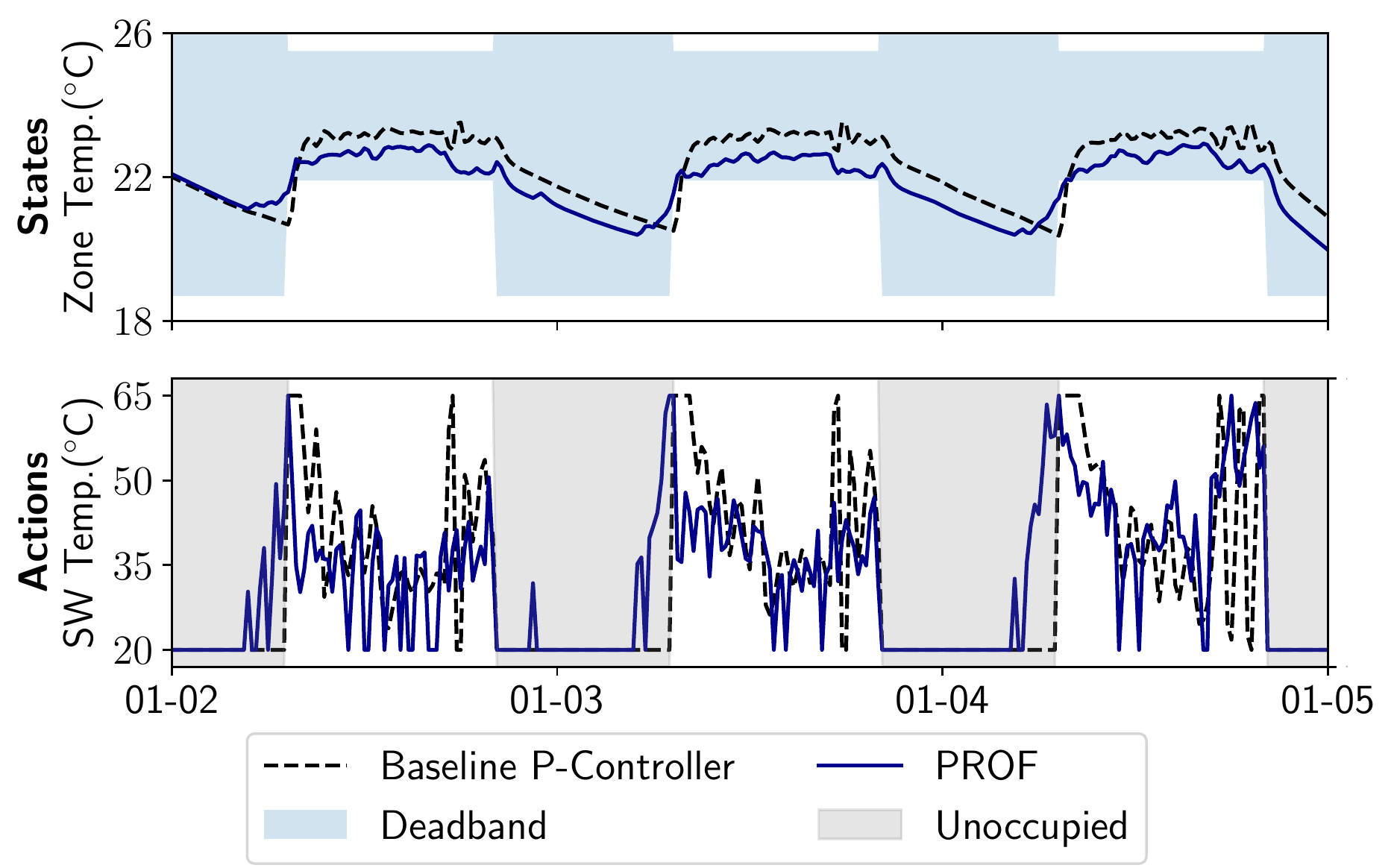}
	\caption{The differentiable projection layer enforces preheating behavior to ensure deadband constraints are never violated, even though this behavior is not present in the expert demonstrations.
	}
	\label{fig:exp1_a}
	\end{subfigure}
	\begin{subfigure}{\linewidth} 
	\includegraphics[width=\textwidth]{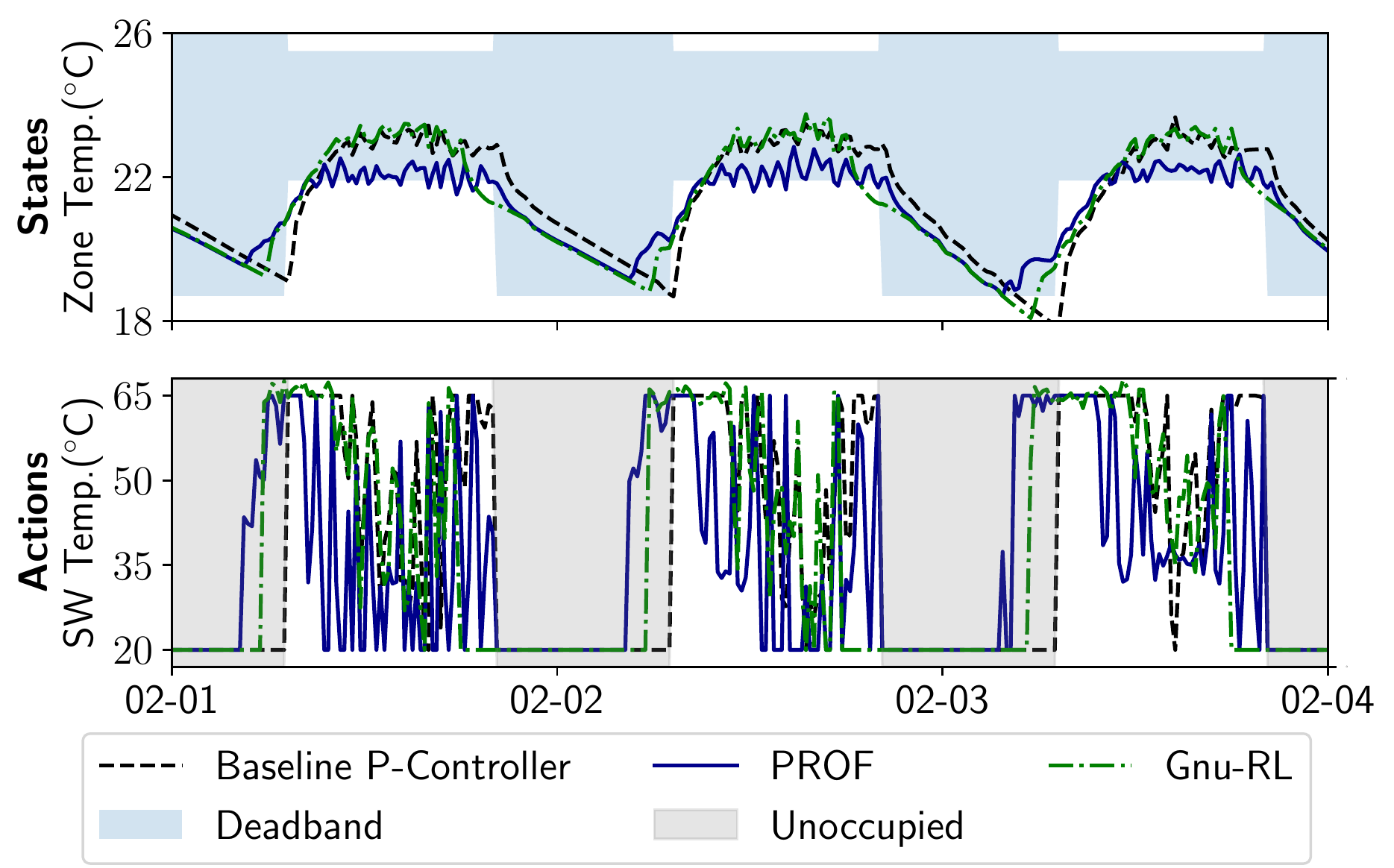}
	\caption{The agent has found a more energy-efficient control strategy by maintaining temperature at the lower end of the deadband.} 
	\label{fig:exp1_b}
	\end{subfigure}
	\caption{Behavior of our proposed agent (a) at the onset of deployment, with pre-trained weights based on expert demonstrations and (b) after a month of interacting with and training on the environment. 
	} \label{fig:exp1}
\end{figure}

After pre-training on expert demonstrations from the baseline P-controller, our agent directly operated the simulation testbed based on actual weather sequences in Pittsburgh from Jan. 1 to Mar. 31 in 2017. Figure \ref{fig:exp1_a} shows the behavior of our agent at the onset of deployment over a 3-day period. The baseline P-controller reactively turns on heating when the environment switches from unoccupied to occupied, which results in thermal comfort violations in the mornings. In comparison, \oursystem~preheats the environment such that the environment is already at a comfortable temperature when occupants arrive in the morning. Notably, the differentiable projection layer manages to enforce this preheating behavior despite this behavior not being present in the expert demonstrations.

Figure \ref{fig:exp1_b} shows the behavior of our agent in comparison with Gnu-RL \cite{chen2019gnu}, having interacted with and trained on the environment for a month. Gnu-RL  is updated via PPO, similarly to the current work, and incorporates domain knowledge on system dynamics. In comparison to Gnu-RL \cite{chen2019gnu}, which ends up trying to maintain temperature at the setpoint, \oursystem ~learns an energy-saving behavior by maintaining the temperature at the lower end of the deadband. This explains the further energy savings compared with Gnu-RL \cite{chen2019gnu}. However, we also notice that the temperature requirement may be violated on cold mornings. This happens when the control action is saturated, i.e., full heating over the 3-hour planning horizon is not sufficient to bring temperature back to the comfortable range. 
(In principle, even these constraint violations could be mitigated by increasing the length of the planning horizon.)

Table \ref{tab:exp_1} summarizes the performance of our agent with comparison to the RL agents in \cite{zhang_buildsys2018, chen2019gnu}. Our proposed agent (averaged over 5 random seeds) saves 10\% and 4\% energy compared to the best-performing agents in \cite{zhang_buildsys2018} and  \cite{chen2019gnu}, respectively. 

\begin{table}[t]
\caption{Performance comparison. Our method saves energy while incurring minimal comfort violations. }
\begin{tabular}{l c c c }\hline 
 &\textbf{Heating}& \multicolumn{2}{c}{ \textbf{PPD}}  \\
 &\textbf{Demand}&\textbf{ Mean}&\textbf{SD}\\
 &\textbf{(kW)}& \textbf{(\%)} & \textbf{(\%)} \\  \hline
Existing P-Controller \cite{zhang_buildsys2018}&43709& 9.45 &5.59 \\
Agent \#6 \cite{zhang_buildsys2018}&37131&11.71&3.76 \\\hline
Baseline P-Controller \cite{chen2019gnu} &35792 &9.71&6.87\\
Gnu-RL \cite{chen2019gnu} &34687
 &9.56 &6.39 \\ \hline
LSTM \& Clip + No Update &37938
 &\textbf{8.55} &3.39 \\
 LSTM \& Clip  &36068 $\pm$ 2187
 &9.18 $\pm$ 0.67 & 3.49 \\
\textbf{\oursystem} (ours)
 &\textbf{33271} $\pm$ 1862
 &9.68 $\pm$ 0.48 & 3.66 \\\hline
\end{tabular}
 \label{tab:exp_1}
\end{table}

We also compare our method to two ablations: (1) \texttt{LSTM \& Clip + No Update}, which uses the same pre-trained weights and the projection layer to enforce feasible actions, but does not update the policy, and (2) \texttt{LSTM \& Clip}, which uses the same pre-trained weights and the projection layer to enforce feasible actions during inference, but does not propagate gradients through the differentiable projection layer in the policy updates. We find that \texttt{LSTM \& Clip} slightly improves upon \texttt{LSTM \& Clip + No Update}, but is less performant compared to \oursystem. This affirms  our hypothesis that the gradients through the differentiable projection layer are cognizant of the constraints and are thus conducive to policy learning.

\section{Experiment 2: Inverter Control}\label{sec:exp-2}

Distributed energy resources (DERs), e.g., solar photovoltaic (PV) panels and energy storage, are becoming increasingly prevalent in an effort to curb carbon dioxide emissions and combat climate change. However, DERs interfacing with the power grid via power electronics, such as inverters, also introduce unintended challenges for grid operators. For instance, over-voltages have become a common occurrence in areas with high renewable penetration \cite{9228929}, and power electronics-interfaced generation has low-inertia and requires active control at much faster timescales compared to traditional synchronous machines \cite{milano2018foundations}. 

To alleviate these issues,  IEEE standard 1547.8-2018 \cite{IEEE1547} recommends a Volt/Var control strategy in which the reactive power contribution of an inverter is based on local voltage measurements. As will be clear in our empirical evaluation, this network-agnostic heuristic based on local information alleviates, but does not avoid, over-voltage issues. Given that the optimal solution needs to be obtained at the system-level and that the problem needs to be solved at very short timescales, a common paradigm is to address the problem in a quasi-static fashion \cite{jalali2019designing} adopted in works such as \cite{baker2017network, jalali2019designing, gupta2020deep}, where one chooses a policy over the next time period, e.g., 15 minutes-1 hour, and uses the policy without update for fast inference. In this work, we adopt the same paradigm and consider real-time control on a 1-second timescale of both active (P) and reactive (Q) power setpoints at each inverter. 


We envision that a neural policy can learn from its prior experiences, in contrast to the traditional \textit{fit-and-forget} approach \cite{dobbe2020learning}, and is capable of making decisions faster compared to solving optimization problems. Our primary contribution compared to existing work is the ability to enforce physical constraints within the neural network. In fact, we successfully enforce voltage constraints 100\% of the time with a randomly initialized neural network, over more than half a million time-steps (i.e., 1 week with a one-second time step). The assumed control and communication scheme is consistent with the new definitions for smart inverter capabilities under IEEE standard 1547.1-2020~\cite{IEEE15472020}.


\subsection{Problem Description} \label{sec:inverter_description}

The problem we are considering here is to control active and reactive power setpoints at each inverter in order to maximize  utilization (i.e., minimize curtailment) of renewable generation, while satisfying the maximum and minimum grid voltage requirements. Here, we first define the considered test case and input data, and describe the model of the network. We refer readers to \cite{baker2017network} for more details on the problem set-up.

\begin{figure}
    \centering
    \includegraphics[width = \linewidth]{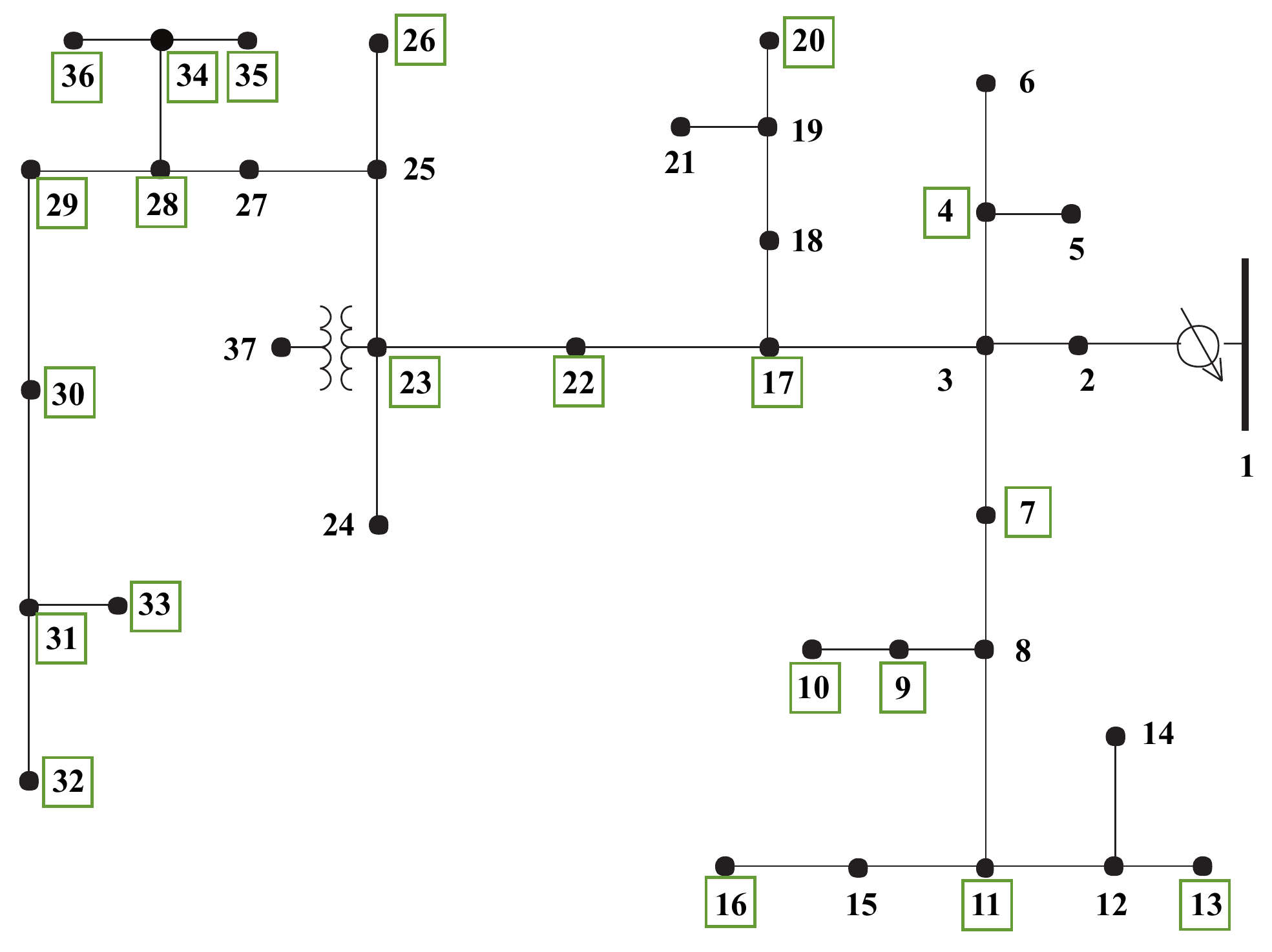}
    \caption{IEEE 37-bus feeder system, where the solar PV systems are indicated by green rectangles.}
    \label{fig:37node}
\end{figure}

\paragraph{IEEE 37-bus test case} We evaluate our method on the IEEE 37-bus distribution test system \cite{37node}, with 21 solar PV systems indicated by green rectangles in Figure \ref{fig:37node}. We utilize a balanced, single-phase equivalent of the system, and simulate the nonlinear AC power flows using PYPOWER \cite{matpower}. For the simulation, the solar generation and loads are based on 1-second solar irradiance and load data collected from a neighborhood in Rancho Cordova, CA \cite{Bank13} over a period of one week (604800 samples).
 
\paragraph{Approximate system model.} Denote the number of buses, excluding the slack bus (e.g., the distribution substation), as $N$, the net active and the reactive power as $\mathbf{p}\in \mathbb{R}^N$ and $\mathbf{q}\in \mathbb{R}^N$, and the voltage at all buses as $\mathbf{v}\in \mathbb{R}^N$. We linearize the AC power flow equations around the flat voltage solution, i.e. $\mathbf{\bar{v}}= \mathbbm{1}$,  using the method in \cite{bolognani2015fast}. The reference active and reactive power corresponding to $\mathbf{\bar{v}}= \mathbbm{1}$ is denoted as  $\mathbf{\bar{p}}$ and $\mathbf{\bar{q}}$. The linearized grid model, $\hat{f}$, is given by Equation~\ref{eq:linear_pf}, where $\mathbf{R},\; \mathbf{B}\in \mathbb{R}^{N\times N}$ represent system-dependent network parameters that can be either estimated from linearization (e.g., \cite{bolognani2015fast}) or data-driven methods:
\begin{equation}
\label{eq:linear_pf}
 \begin{aligned}
        \mathbf{v} \approx \hat{f}(\mathbf{p}, \mathbf{q}) &= \mathbf{\bar{v}}  
               + \mathbf{R}(\mathbf{p-\bar{p}})+\mathbf{B}(\mathbf{q-\bar{q}})\\
        &= \mathbf{\bar{v}} + \underbrace{[\mathbf{R}, \mathbf{B}]}_{\mathbf{H}}\underbrace{\begin{bmatrix}
               \mathbf{p-\bar{p}}, \\
               \mathbf{q-\bar{q}}
               \end{bmatrix}
               }_{\mathbf{u}}.
    \end{aligned}
\end{equation}
A notable advantage of the method in \cite{bolognani2015fast} is that the resulting model has bounded error with respect to the true dynamics. By incorporating the error bound when constructing the safety set, the safety set is guaranteed to be a conservative under-approximation of the true safety set, and thus allow us to satisfy voltage constraints 100\% of the time. 

\paragraph{Policy} Our policy takes as input the voltage from the previous time-step, load, and generation at all the buses, and outputs active and reactive power setpoints at each inverter.
(This is a deterministic policy; see Equation~\ref{eq:prob-det}.)
Note that while the grid model (Equation~\ref{eq:linear_pf}) contains all $N$ buses, only those with inverters are controllable. 

Our neural architecture is similar to the one used in \cite{gupta2020deep}, which consists of a utility-level network, and inverter-level networks for individual inverters. The utility-level network collects information from all nodes, and broadcasts an intermediate representation to all inverter-level networks. Using this information along with its local observations, each inverter makes its local control decisions, which are then projected onto the constraints (discussed below).
\begin{figure*}
    \centering
    \includegraphics[width = \linewidth]{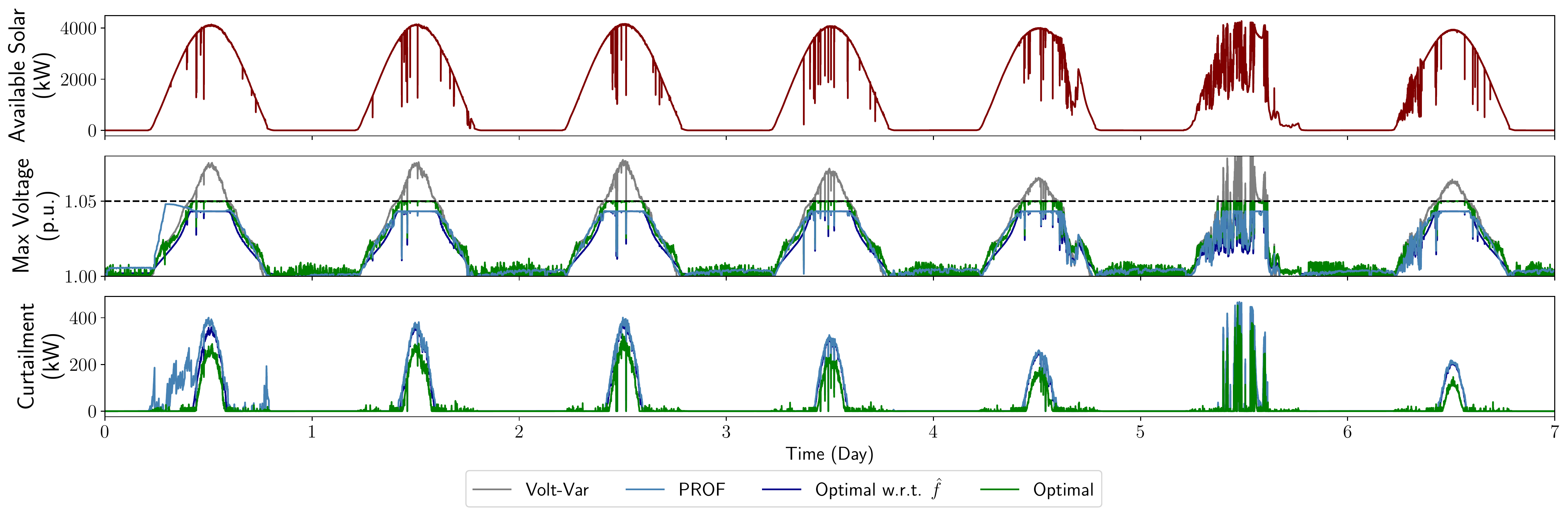}
    \caption{\oursystem~satisfies voltage constraints throughout the experiment, and learns to minimize curtailment as well as possible within its conservative safety set, $\hat{C}_k$, after learning safely for a day.}
    \label{fig:exp_2}
\end{figure*}
\paragraph{Objective} The objective is to minimize the curtailment of solar generation, or equivalently to maximize the utilization of the available solar power, $p_{av}$.
Specifically, letting $\mathcal{I}$ denote the set of buses with inverters, the objective is

\begin{equation}\label{eq:solar_curtail}
J(\theta) = \min_{\mathbf{p}_{\mathcal{I}}, \mathbf{q}_{\mathcal{I}}} \sum_{i \in \mathcal{I}} [p_{av,i} - p_i]_+,\quad \text{where} \begin{bmatrix} \mathbf{p}_{\mathcal{I}} &  \mathbf{q}_{\mathcal{I}} \end{bmatrix} =\pi_\theta
\end{equation}

\paragraph{Constraints} For an individual inverter, $i$, with rated power $s_i$ and an available power (from available solar generation) $p_{av, i}$, the feasible action space is
$$\mathcal{U}_i (k) = \{ (p_i, q_i) : 0\leq p_i \leq p_{av, i}(k), p^2_i + q^2_i \leq s^2_i \}$$  
$$\mathcal{U} (k) := \mathcal{U}_1 (k) \times\dots \times \mathcal{U}_{|\mathcal{I}|} (k). $$
%
At the same time, the voltage at each bus should remain between $0.95$-$1.05~p.u.$ The primary challenge of satisfying voltage constraints is that the voltage at each bus depends on actions of neighboring nodes, i.e. 
$$\mathcal{X} = \{v\;|\;0.95\times \mathbbm{1} \leq \mathbf{v}\approx \mathbf{\bar{v}}+ \mathbf{H} \mathbf{u} \leq 1.05\times \mathbbm{1}\},$$
where the sparsity pattern of $H$ is characterized by the admittance matrix.
We jointly project actions from all inverters at each time step $k$ onto the constraints $\mathcal{U}(k) \cap \mathcal{X}$.


\subsection{Implementation Details}
We evaluate \oursystem~by executing it over the 1-week dataset (at 1 second) once. Similarly to other quasi-static approaches, we update the policy every 15-minutes. Similarly to \cite{gupta2020deep}, we optimize the neural policy with stochastic samples by directly differentiating through the objective (Equation \ref{eq:solar_curtail})  and   the linearized grid model (Equation \ref{eq:linear_pf}). However our method differs in that \citet{gupta2020deep} characterized the constraints as a regularization term, and learned the policy via primal-dual updates. We  incorporate the constraints directly via the differentiable projection layer and thus guarantee constraint satisfaction. 

We use $\lambda$=10 (see Equation~\ref{eq:aug-cost}), a learning rate of $10^{-3}$,  and RMSprop \cite{tieleman2012lecture} as the optimizer. At every 15 minutes, we sample 16 batches of data with size of 64 from the replay memory. We keep a replay memory size of 86400, i.e., samples from the previous day. For the both the utility-level network and the inverter-level network, we use fully-connected layers with ReLU activations. The utility-level network has hidden layer sizes (256, 128, 64), and each inverter-level network has hidden layer sizes (16, 4) and outputs active and reactive power. On top of the neural network, we implement the differentiable projection layer, following the constraints described in Section \ref{sec:inverter_description}.

We compare our methods to three baselines, (1) a Volt/Var strategy following IEEE 1547.8 \cite{IEEE1547},  (2) the optimal solution with respect to the linearized grid model, and (3) the optimal solution with respect to the true AC power flow equations.

\subsection{Results}
The performance of \oursystem~in comparison to the three baselines is summarized in Figure~\ref{fig:exp_2}. For clarity, we only show the maximum voltage over all buses; under-voltage is not a concern for this particular test case.

We see that the Volt/Var strategy violates voltage constraints 22.3\% of time, mostly around noon when the solar generation is high and there is a surplus of energy. Since the Volt/Var baseline does not adjust active power, there is no curtailment. 

In comparison, \oursystem~satisfies the voltage constraints throughout the experiment, even with a randomly initialized neural policy. While \oursystem~performs poorly on the first morning, it quickly improves its policy. In fact, the behavior of \oursystem~is barely distinguishable from the optimal solution with respect to the linearized grid model, after learning safely for a day. This implies that \oursystem~learned to control inverters as well as possible given its approximate model, which constructs a conservative under-approximation of the true safety set. 

The optimal baseline with respect to the true AC power flow equations unsurprisingly achieves the best performance with respect to minimizing curtailment, as it can 
push the maximum voltage to the allowable limit in order to maximally reduce the amount of curtailed energy. 
However, inverter control is a task that requires near real-time inputs, and we find that running this baseline can be prohibitively slow.
Specifically, 
we evaluate the computation time of different operations by averaging over 1000 randomly sampled problems from our dataset on a personal laptop. 
For \oursystem, on average, a forward pass in the neural network (excluding the projection layer) took 4.5 ms and the differentiable projection operation took 8.6 ms. The computation cost of the differentiable projection could be further reduced by using customized projection solvers such as the ones in \cite{amos2017optnet, donti2021enforcing} that avoid the ``canonicalization'' costs introduced by general-purpose solvers such as the one we use \cite{agrawal2019differentiable}. 
In comparison, solving the optimization baseline with respect to the true AC power flow equations took 1.02s on the same machine, which is even longer than the 1s control time-step. 

\section{Discussion and Conclusions}

In this work, we have presented a method,~\oursystem, for integrating convex operational constraints into neural network policies for  energy systems applications.
In particular, we propose a policy that entails passing the output of a neural network to a differentiable projection layer, which enforces a convex approximation of the operational constraints.
These convex constraint sets are obtained using  approximate models of the system dynamics, which can be fit using system data and/or constructed using domain knowledge.
We can then train the resultant neural policy via standard RL algorithms, using an augmented cost function designed to effect desirable policy gradients.
The result is that our neural policy is cognizant of relevant operational constraints during learning, enhancing overall performance.

We find in both the building energy optimization and inverter control settings that \oursystem~successfully enforces relevant constraints while improving performance on the control objective.
In particular, in the building thermal control setting, we find that our approach achieves a 4\% energy savings over the state of the art while largely maintaining the temperature within the deadband.
In the inverter control setting, our method perfectly satisfies the voltage constraints over more than half a million time steps, while learning to minimize curtailment as much as possible within the safety set. 

While these results demonstrate the promise of our method, a key limitation is in its computational cost.
In particular, computing a projection during every forward pass of training and inference is decidedly more expensive than running a ``standard'' neural network.
A fruitful area for future work -- both in the context of our method, and in the context of research in differentiable optimization layers as a whole -- may be to improve the speed of such differentiable projection layers.
For instance, this might entail developing special-purpose differentiable solvers \cite{amos2017optnet, donti2021enforcing} for optimization problems commonly encountered in energy systems applications, developing approximate solvers that do not rely on obtaining optimal solutions in order to compute reasonable gradients, or employing cheaper projection schemes such as $\alpha$-projection \cite{shah2020solving} where possible.

Additionally, the success of our method (and many other constraint enforcement methods) depends fundamentally on the quality of the approximate model used to characterize the constraint sets.
In particular, this determines the extent to which the resultant approximate constraint sets are a good representation of the true operational constraints.
While we were able to employ reasonably high-quality approximation schemes in the context of this work, future work on safely updating the models or the constraint sets directly \cite{fisac2019bridging} may greatly improve the quality of the solutions.


More generally, while our work highlights one approach to enforcing physical constraints within learning-based methods, we believe this is only the start of a broader conversation on closely integrating domain knowledge and control constraints into learning-based methods. 
In particular, strictly enforcing physical constraints will be paramount to the real-world success of these methods in energy systems contexts, and we hope that our paper will serve to spark further inquiry into this important line of work.


\section{Acknowledgments}
This material is based, in part, on work supported by Carnegie Mellon University’s College of Engineering Moonshot Award for Autonomous Technologies for Livability and Sustainability (ATLAS). This work was also supported by the U.S.~Department of Energy Computational Science Graduate Fellowship (DE-FG02-97ER25308), the Center for Climate and Energy Decision Making through a cooperative agreement between the National Science Foundation and Carnegie Mellon University (SES-00949710), the Computational Sustainability Network, and the Bosch Center for AI. The work of K. Baker is supported by the National Science Foundation CAREER award 2041835. 

\bibliographystyle{ACM-Reference-Format}
\bibliography{main}

\end{document}